\documentclass[fleqn,usenatbib]{mnras}

\usepackage[T1]{fontenc}
\usepackage{ae,aecompl}

\usepackage{amsmath,siunitx,booktabs,threeparttable}
\usepackage{hyperref}
\usepackage{pgfplots}
\pgfplotsset{compat=newest}
\usepgfplotslibrary{external}
\usepackage{graphicx}
\usepackage{xfrac}
\usepackage{commath}

\newcommand{\flash}{\textsc{flash}}
\newcommand{\piernik}{\textsc{piernik}}

\graphicspath{{figures/}
{/home/mdevalbo/FLASH/FLASH2.5/runs/mira-isot-della4/}
{/home/mdevalbo/project/hydro/piernik-dev/runs/wind/frames/}
{/home/mdevalbo/project/hydro/piernik-dev/runs/wind/}
{/home/mdevalbo/project/hydro/runs/piernik/wind/wide/}
{/home/mdevalbo/project/hydro/runs/piernik/wind/wide2/}
}

\makeatletter
\newlength{\abovecaptionskip}%
\makeatother

\DeclareSIUnit\msun{\text{\ensuremath{M_{\odot}}}}
\DeclareSIUnit\rsun{\text{\ensuremath{R_{\odot}}}}
\DeclareSIUnit\lsun{\text{\ensuremath{L_{\odot}}}}
\DeclareSIUnit\year{yr}

\defcitealias{2009ApJ...700.1148D}{Paper~I}

\title[3D models of wind accretion in binaries]{Three-dimensional hydrodynamical models of wind and
outburst-related accretion in symbiotic systems}

\author[M.~de~Val-Borro, M.~Karovska, D.~D.~Sasselov and J.~M.~Stone]%
{M.~de~Val-Borro,$^{1,2}$\thanks{E-mail:
\href{mailto:miguel.devalborro@nasa.gov}{miguel.devalborro@nasa.gov} (MdVB);
\href{mailto:mkarovska@cfa.harvard.edu}{mkarovska@cfa.harvard.edu} (MK)}
M.~Karovska,$^{3}$\footnotemark[1]
D.~D.~Sasselov$^{3}$
and J.~M.~Stone$^{4}$
\\
$^{1}$NASA Goddard Space Flight Center, Astrochemistry Laboratory, 8800 Greenbelt Road, Greenbelt, MD 20771, USA\\
$^{2}$Department of Physics, Catholic University of America, Washington,
DC 20064, USA\\
$^{3}$Harvard-Smithsonian Center for Astrophysics, 60 Garden Street, Cambridge, MA
02138, USA\\
$^{4}$Department of Astrophysical Sciences, Princeton University, Princeton,
NJ 08544, USA
}

\date{Accepted 2017 March 15.
Received 2017 March 15;
in original form 2016 December 9}

\pubyear{2017}

\begin{document}
\label{firstpage}
\pagerange{\pageref{firstpage}--\pageref{lastpage}}
\maketitle

\begin{abstract}
Gravitationally focused wind accretion in binary systems consisting of
an evolved star with a gaseous envelope and a compact accreting
companion is a possible mechanism to explain mass transfer in symbiotic
binaries.  We study the mass accretion around the secondary caused by
the strong wind from the primary late-type component using global
three-dimensional hydrodynamic numerical simulations during quiescence
and outburst stages.  In particular, the dependence of the mass
accretion rate of the  mass-loss rate, wind parameters and phases of
wind outburst development is considered.  For a typical wind from an
asymptotic giant branch star with a mass-loss rate of
\SI{e-6}{\msun\per\year} and wind speeds of \SIrange{20}{50}{\km\per\s},
the mass transfer through a focused wind results in efficient infall on
to the secondary.  Accretion rates on to the secondary of 5--20 per cent
of the mass-loss from the primary are obtained during quiescence and
outburst periods where the wind velocity and mass-loss rates are varied,
about 20--50 per cent larger than in the standard
Bondi--Hoyle--Lyttleton approximation.  This mechanism could be an
important method for explaining observed accretion luminosities and
periodic modulations in the accretion rates for a broad range of
interacting binary systems.
\end{abstract}

\begin{keywords}
Accretion, accretion discs
-- Binaries: symbiotic
-- Circumstellar matter
-- Methods: numerical
-- Stars: mass-loss
\end{keywords}

\section{Introduction}
\label{sec:intro}

Symbiotic binaries are active bright systems with a composite spectrum
that exhibits absorption features together with strong H and He emission
lines.  These binaries are important astrophysical laboratories for
studies of wind accretion because of the wide separation of the
components, and the ability to study the individual components and the
accretion processes in close circumbinary environments at multiple
wavelengths ranging from X-ray to radio
\citep{1997ApJ...482L.175K,2005ApJ...623L.137K,2010ApJ...710L.132K}.  A
typical symbiotic system consists of a mass-losing red giant or
asymptotic giant branch (AGB) star and a hotter accreting companion,
often a white dwarf (WD).  The components in these systems are believed
to be detached (both components are well within their Roche lobes),  and
the activity is caused by the accretion of mass from the massive wind of
the cool evolved star on to the compact companion at rates determined by
the wind and orbital parameters \citep[e.g.,][]{1986syst.book.....K}.

Mass-loss is known to be a key factor in the late stages of the
evolution on the red giant branch (RGB) and on the AGB and regulates how
the star evolves.  Matter escapes easily because of the low surface
gravity and the stellar wind removes angular momentum from the star.
Accretion of this material can occur in two different modes: Roche lobe
overflow (RLOF) when the primary star fills its Roche surface \citep[see
e.g.,][]{1971ARA&A...9..183P} or wind accretion
\citep[e.g.,][]{1993MNRAS.265..946T}.  Considerable differences are seen
between simulations contingent upon whether the dust acceleration radius
is close to the Roche lobe surface or not (i.e., dependent on the binary
separation).  In RLOF accretion, the donor star fills its Roche lobe and
mass transfer occurs through the Lagrangian point $L_1$ on to the
companion's Roche lobe.  Accretion in the canonical RLOF scenario is
quasi-conservative, with the mass accretion rate on to the secondary
being very similar to the mass-loss rate from the primary.  None the less
some material may escape the binary system through the $L_2$ point so
the mass rate ratio $\dot{M}_\mathrm{acc}/\dot{M_1}$ may be smaller than 1.

In the quiescent phase of symbiotic binaries the fraction of the wind
that is accreted by the hot companion is ionized and powers the observed
luminosity in these systems. However, high accretion rates of the order
of \SIrange{e-7}{e-6}{\msun\per\year} are required to maintain the high
luminosity of symbiotic systems, which is about an order of magnitude
larger than the values derived from standard Bondi--Hoyle--Lyttleton
wind accretion \citep[BHL;][]{1939PCPS...35..405H,1944MNRAS.104..273B}.
To solve this problem, a number of numerical studies of a
gravitationally focused wind in Mira-type systems have been carried out
recently, which produced an enhanced mass transfer in binaries compared with BHL
accretion
\citep[e.g.,][]{2004A&A...419..335N,2007BaltA..16...26P,2009ApJ...700.1148D}
and are capable of explaining the observed mass outflow geometries in
some symbiotic systems
\citep[e.g.,][]{2005ApJ...623L.137K,2010ApJ...710L.132K,2011AAS...21822803K}.

Previous two-dimensional models of wind accretion in detached binaries
develop complex flow patterns and the formation of a stream flow between
the components, which is dependent on the wind and orbital parameters of
the system.  These binary models suggest that the accretion rate on to
the secondary is considerably modified by the effect of the
gravitational interaction with a companion star
\citep[e.g.,][]{2009ApJ...700.1148D}.  For slow radiative driven winds,
the accretion efficiency on the companion is of the order of 10--20
per cent mass-loss rate of the primary and can be increased compared with the
standard BHL accretion in two- or three-dimensional simulations
\citep[e.g.,][]{2009ApJ...700.1148D,2012BaltA..21...88M}.

An additional possible mode to obtain a focused wind in the orbital
plane in S-type symbiotic systems is the wind compression disc
model that can enhance the efficiency of wind accretion
\citep{1993ApJ...409..429B,2015A&A...573A...8S}.  Wind accretion can
also lead to a higher enrichment of barium and carbon in metal-poor
stars and changes in the orbital parameters because of angular momentum
loss from the binary system \citep[e.g.,][]{2013A&A...552A..26A}.
Nonetheless, the applicability of this mode of accretion to large
separation symbiotic binaries remains to be established precisely.

Numerical simulations of BHL accretion in a two-dimensional planar flow
have shown the presence of instabilities in the accretion shock,
so-called flip-flop instabilities
\citep{1987MNRAS.226..785M,1988ApJ...335..862F}, which have been suggested
to occur in several systems. In this instability, the accretion flow
oscillates between states with opposite spin with a brief intermediate
phase.  Some analytical approaches have been proposed using a
perturbation analysis on a stationary flow solution \citep[see
e.g.,][]{1991MNRAS.253..633L}. This instability may cause periodic
oscillations that can explain some astrophysical phenomena such as
fluctuations in X-ray binaries and pulsating systems
\citep{1988ApJ...335..862F}. However, three-dimensional simulations have
found that the instability is not as strong as in the two-dimensional
case, or may be completely absent
\citep[e.g.,][]{1994ApJ...427..351R,2012ApJ...752...30B}.

There is observational evidence of small-scale sporadic outbursts and
stochastic variations on short time-scales \citep[as seen in the light
curves, spectra and images of symbiotic
systems;][]{1978MNRAS.185..591S,1986Natur.319...38T}, which could be due
to a change of the accretion rate and the accretion environment on
time-scales much smaller than the orbital period. The origin of these
outbursts is still a puzzle. It is possible that they are caused by a
temporary increase of mass accretion on to the companion due to changes
in mass-loss and wind velocity in the red giant component or a
fluctuating accretion disc around the secondary.  Gas shells produced by
the variable mass-loss from AGB stars have been observed in several
systems since the first detection by \citet{1990A&A...230L..13O}.  A
shell around pulsating star R Sculptoris has been clearly resolved with
the Atacama Large Millimeter/submillimeter Array showing a spiral
structure that is formed by interaction with a previously unknown
companion \citep{2012Natur.490..232M,2016A&A...586A...5M}.  Recently,
\citet{2016MNRAS.457..822B} have simulated the formation of asymmetric
circumstellar shells using a smoothed particle hydrodynamics (SPH) code.
Their model produces a bipolar structure surrounding the recurrent nova
progenitor in the RS Ophiuchi symbiotic binary, a system known to
undergo recurrent outbursts that have been observed at multiple
wavelengths \citep{2006Natur.442..279O,2007ApJ...665L..63B}.

Here we present three-dimensional hydrodynamical simulations of
gravitationally focused wind accretion in symbiotic binary systems using
as an example the nearby jet-ejecting CH Cygni (CH Cyg) system.  This is
a variable symbiotic that is of special interest to examine the basic
physical processes of mass transfer due to its brightness and close
distance, $d \sim \SI{250}{pc}$, to Earth.  We address the variability
in this system by studying stellar outbursts in the primary to consider
the effect in the outflow properties and accretion rates.  In a future
work, we will carry out a detailed comparison of observational
diagnostics in the symbiotic variable CH Cyg system with the results of
our wind accretion simulations.

In Section~\ref{sec:model}, we describe the numerical set-up.  In
Section~\ref{sec:results}, we show our results from hydrodynamical
simulations of the symbiotic binary CH Cyg, and discuss the flow
structure  around the accretor, accreting rates, and the circumbinary
environment.  Finally, we summarize the main results of this work in
Section~\ref{sec:discuss}.

\section{Numerical Model}
\label{sec:model}

Realistic models of mass transfer in symbiotic binaries
require computationally demanding hydrodynamical simulations.
The wind gas is modelled using the basic Euler equations of
hydrodynamics describing the evolution of the density and velocity field
in three dimensions \citep[see e.g.,][]{2014pafd.book.....C}:
\begin{align}
  \frac{\partial\rho}{\partial t} + \nabla \cdot (\rho \mathbf{v}) & = 0,\\
  \frac{\partial\mathbf{v}}{\partial t} + (\mathbf{v} \cdot \nabla )
  \mathbf{v} & = - \frac{1}{\rho}\nabla P - \nabla \Phi.
  \label{eq:NS}
\end{align}
In the above equations, symbols have their usual meaning: $\rho$ is the
density of the fluid, $\mathbf{v}$ is the velocity, $P$ is the pressure and
$\Phi$ is the stellar gravitational potential that is given by the point
mass equation with softening length $\epsilon$ to simplify the
numerical integration of the equations.
\[
\Phi = \Phi_\mathrm{1} +\Phi_\mathrm{2} = -
\frac{G M_\mathrm{1}}{\sqrt{|\mathbf{r} -
\mathbf{r}_\mathrm{1}|^2 + \epsilon_0^2}}
- \frac{G M_\mathrm{2}}{\sqrt{|\mathbf{r} - \mathbf{r}_\mathrm{2}|^2 +
\epsilon_0^2}},
\]
where $M_\mathrm{1}$ and $M_\mathrm{2}$ are the stellar masses of the
two components, and $\mathbf{r}_1$ and $\mathbf{r}_2$ are their positions.
The softening length is chosen to be $\epsilon_0 =
\SI{0.05}{au}$.  We note that the self-gravity of the fluid is not
considered.  The wind pressure in the fluid is derived using the
polytropic equation of state for an ideal gas:
\[
  P = (\gamma - 1)\rho\epsilon,
\]
where $\epsilon$ is the internal energy per unit mass and $\gamma$ is the
adiabatic index of the gas, the ratio of specific heat at constant
pressure and volume, which has a value of $\gamma = \sfrac{5}{3}$ for the
monoatomic gas in our model. Thus, the value of the polytropic index
is given by $n = (\gamma-1)^{-1} = \sfrac{3}{2}$.
We set the mean atomic number to be the
atomic hydrogen weight $\mu = 1$.
The sound speed of the fluid for an ideal gas law is described by:
\[
  c_s = \sqrt{\frac{\gamma P}{\rho}}.
\]
We use the Courant--Friedrichs--Lewy condition to constrain the time-step
to ensure the numerical stability of the solution.

The numerical model is similar to previous two-dimensional models
described in \citet{2009ApJ...700.1148D}, hereafter referred to as
\citetalias{2009ApJ...700.1148D}. However, the model uses a different
code base with the equation of state for a polytropic fluid
and the equations of hydrodynamics are solved in three dimensions.
We take into account the rotation of the binary and assume circular
Keplerian orbits for the system with the stellar components at rest
to avoid stability issues.
We work in the non-inertial reference frame that rotates around the
origin with a constant angular frequency $\Omega$, which is given by the
equation
\[
  \Omega = \sqrt{\frac{G(M_\mathrm{1} + M_\mathrm{2})}{a^3}},
\]
where $a$ is the semimajor axis of the system.  The two stellar
components are treated as point masses with a smoothing radius in our
models.  The secondary is treated as an absorbing sphere, and our model
includes a smooth variation of the wind properties during a short
fraction of an orbital period that translates into changes in the
accretion rates on to the companion.  We do not resolve the surface of
the secondary in our calculation.

\subsection{Stellar wind}
\label{sec:cartesian}

\begin{figure}
  \centering
  \includegraphics[width=\columnwidth]{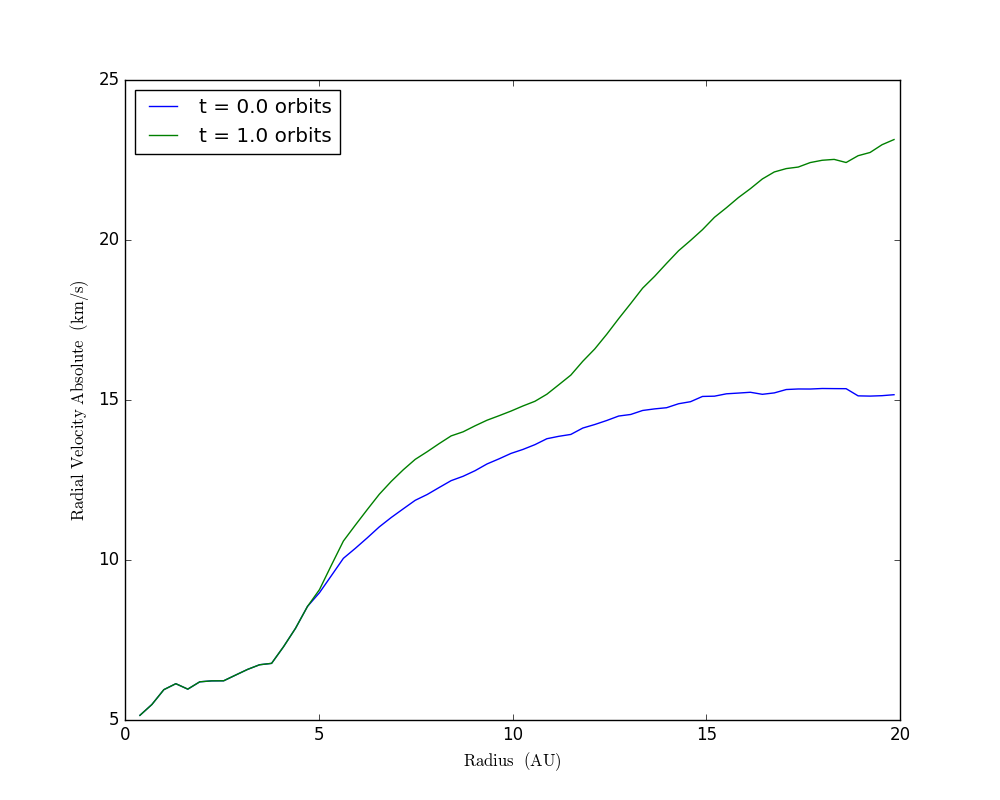}
  \caption{Averaged radial velocity as a function of radius centred on
  the location of the mass-losing star at two different times. The
  secondary is located at a distance of \SI{10}{au}.
  }
  \label{fig:vel_prof}
\end{figure}

We have developed a single-fluid stellar wind model implemented in a
Cartesian grid in a rotating frame of reference where both stars are at
rest in the computational domain based on the publicly available
\piernik{} magnetohydrodynamics code
\citep{2010EAS....42..275H}\footnote{The \piernik{} source code is
hosted in a public Git repository at \url{https://github.com/piernik-dev/piernik} under the
GNU General Public License v3.}.  The numerical algorithm is based on
the original relaxing total variation diminishing (RTVD) method, which is
a second-order algorithm in space and time \citep[][and references
therein]{Jin95,2003ApJS..149..447P}. One of the advantages of this
method is that it can handle shocks and discontinuities in the flow with
high resolution without using artificial viscosity.  In our calculations
the wind material is treated using the equation of state of a classical
ideal gas.  The effect of the Coriolis and the centrifugal forces
has been included in the equations for each sweep in a low-order
approximation using the gas density and velocity values in the beginning
of the time-step in an operator-split way.  Strict numerical
conservation of angular momentum to numerical precision is not achieved
as we use a Cartesian coordinate grid with open outflow boundaries.
Spurious changes in angular momentum occur at the outer boundary of the
grid.  Other solvers are found to be less diffusive than the RTVD
method, and using cylindrical coordinates can bring substantial improvement in
the conservation of angular momentum, in particular, at the outer
boundary of the domain
\citep[e.g.,][]{2008ApJS..178..137S,2013MNRAS.434.1460K}.  However, we
have chosen a large enough computational domain that these changes are
not expected to affect the accretion process.  Further details of the
used numerical scheme can be found in \citet{2003ApJS..149..447P}.

\piernik{} uses an adaptive mesh refinement (AMR) implementation
that adapts the resolution dynamically with a criterion that can be
defined by the user.
The AMR method allows us to efficiently study binary
systems with variable mass-loss, where a perturbation propagates to large
radii by refining cells at locations with sharp contrasts in density.
We used the standard grid refinement criterion based on the second
derivative of the gas density and pressure that has been used in the
study of the tidal interaction between a protoplanetary disc and an
embedded planet \citep[see
e.g.][]{2006MNRAS.370..529D,2008PhDT.......201D}.  For our simulations,
we use an adaptive grid with three to five additional refinement levels.

The code is fully parallelized using a block-structured decomposition
that runs on any platform that supports the Message Passing Interface,
and it has been found to scale well up to \num{1000} CPU cores
\citep{2010EAS....42..275H}.  \piernik{} can be extended by adding new
physics modules and has been extensively tested in various compressible
flow problems such as simulations of the streaming instability in
protoplanetary discs \citep{2013MNRAS.434.1460K}.

The stellar wind is characterized by its mass-loss $\dot{M}$, terminal
wind velocity $v_\infty$ and dust acceleration radius. Mass-loss in
evolved stars is driven by the uplift of gas molecules due to pulsation
shocks that condense into dust grains at a certain distance from the
star where they are accelerated by radiative pressure.  An unperturbed
initial velocity profile is given by the simplified analytical solution
of a steady-state wind that considers the effect of radiation pressure
in the mass-losing star and the pressure gradient
\citep{1958ApJ...128..664P,1963idp..book.....P}.  We impose outflow
boundary conditions at all boundaries in the computational domain and
constant density in the region inside the wind acceleration radius.
Fig.~\ref{fig:vel_prof} shows the azimuthally averaged radial velocity
in the wind at the beginning of the simulation and after one orbit.

\begin{figure*}
  \centering
  \begin{tikzpicture}
    \begin{axis}[
      enlargelimits=false,
      axis equal image,
      axis on top,
      ylabel={$z$ [au]},
      name=xz]
      \addplot graphics
         [xmin=-14.875,xmax=14.875,ymin=-14.875,ymax=14.875]
         {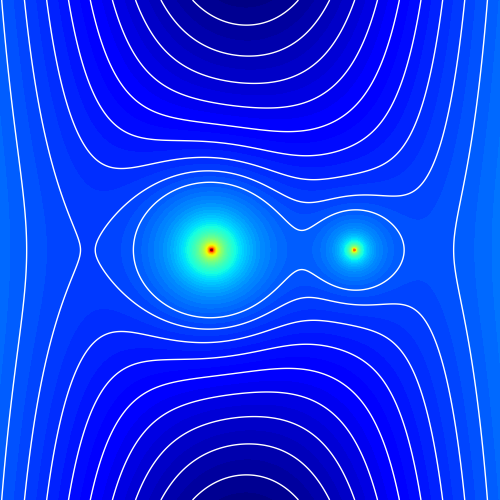};
      \addplot[mark=+,color=white] coordinates {(3.32,0)}
	node [above] {$L_1$};
      \addplot[mark=+,color=white] coordinates {(10.78,0)}
	node [above] {$L_2$};
      \addplot[mark=+,color=white] coordinates {(-9.8,0)}
	node [above] {$L_3$};
      \addplot[mark=+,color=white] coordinates {(0,0)}
	node [above] {$C$};
    \end{axis}
    \begin{axis}[
      enlargelimits=false,
      axis equal image,
      axis on top,
      xlabel={$x$ [au]},
      ylabel={$y$ [au]},
      at=(xz.south), anchor=north]
      \addplot graphics
         [xmin=-14.875,xmax=14.875,ymin=-14.875,ymax=14.875]
         {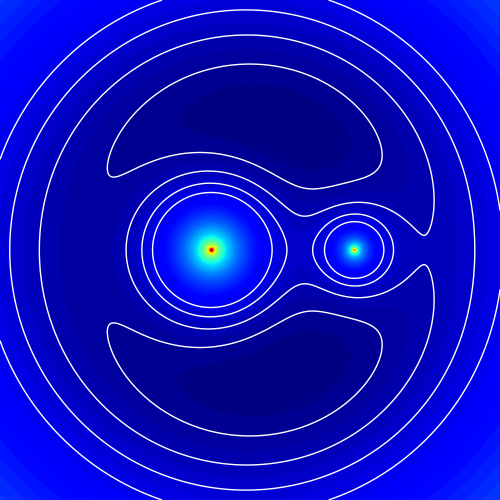};
      \addplot[mark=+,color=white] coordinates {(3.32,0)}
	node [above] {$L_1$};
      \addplot[mark=+,color=white] coordinates {(10.78,0)}
	node [above] {$L_2$};
      \addplot[mark=+,color=white] coordinates {(-9.8,0)}
	node [above] {$L_3$};
      \addplot[mark=+,color=white] coordinates {(2.29,7.36)}
	node [above] {$L_4$};
      \addplot[mark=+,color=white] coordinates {(2.29,-7.36)}
	node [above] {$L_5$};
      \addplot[mark=+,color=white] coordinates {(0,0)}
	node [above] {$C$};
    \end{axis}
    \begin{axis}[
      enlargelimits=false,
      axis equal image,
      axis on top,
      at=(xz.east), anchor=west,
      yticklabels={,,},
      name=xz2]
      \addplot graphics
         [xmin=-14.875,xmax=14.875,ymin=-14.875,ymax=14.875]
         {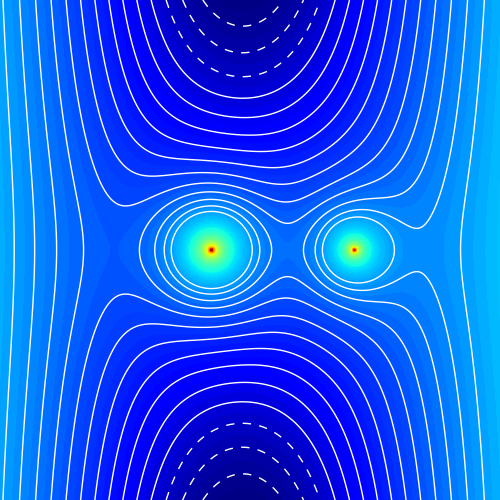};
      \addplot[mark=+,color=white] coordinates {(2.32,0)}
	node [above] {$L_1$};
      \addplot[mark=+,color=white] coordinates {(10,0)}
	node [above] {$L_2$};
      \addplot[mark=+,color=white] coordinates {(-8,0)}
	node [above] {$L_3$};
      \addplot[mark=+,color=white] coordinates {(0,0)}
	node [above] {$C$};
    \end{axis}
    \begin{axis}[
      enlargelimits=false,
      axis equal image,
      axis on top,
      xlabel={$x$ [au]},
      yticklabels={,,},
      at=(xz2.south), anchor=north]
      \addplot graphics
         [xmin=-14.875,xmax=14.875,ymin=-14.875,ymax=14.875]
         {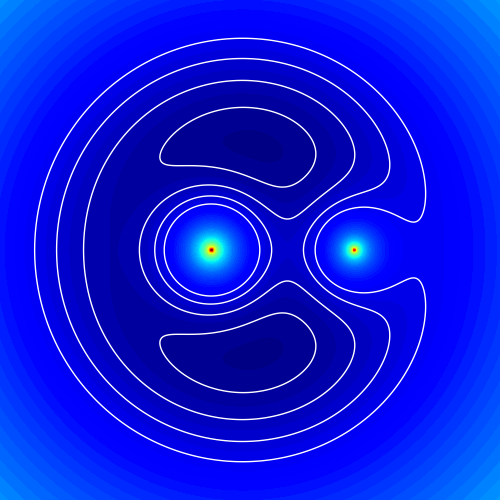};
      \addplot[mark=+,color=white] coordinates {(2.32,0)}
	node [above] {$L_1$};
      \addplot[mark=+,color=white] coordinates {(10,0)}
	node [above] {$L_2$};
      \addplot[mark=+,color=white] coordinates {(-8,0)}
	node [above] {$L_3$};
      \addplot[mark=+,color=white] coordinates {(1, 6)}
	node [above] {$L_4$};
      \addplot[mark=+,color=white] coordinates {(1,-6)}
	node [above] {$L_5$};
      \addplot[mark=+,color=white] coordinates {(0,0)}
	node [above] {$C$};
    \end{axis}
  \end{tikzpicture}
  \caption{Schematic Roche potential contours on the orbital $x$--$y$ plane
  (lower left-hand panel) and perpendicular $x$--$z$ plane (upper left-hand panel) of
  the CH Cyg binary system (mass ratio $q= M_\mathrm{2}/M_\mathrm{1} = 0.3$)
  The $x$--$y$ plane (lower right-hand panel) and perpendicular $x$--$z$ plane (upper
  right-hand panel) show the Roche potential for the same system with a
  ratio of radiation pressure force per unit mass to the
  gravitational force per unit mass $f_\mathrm{grav} = 0.5$.
  The white crosses indicate the centre of mass ($C$) and the
  Lagrangian points ($L_1$--$L_5$) where the gradient of the potential
  is zero.
  }
  \label{fig:chcyg}
\end{figure*}
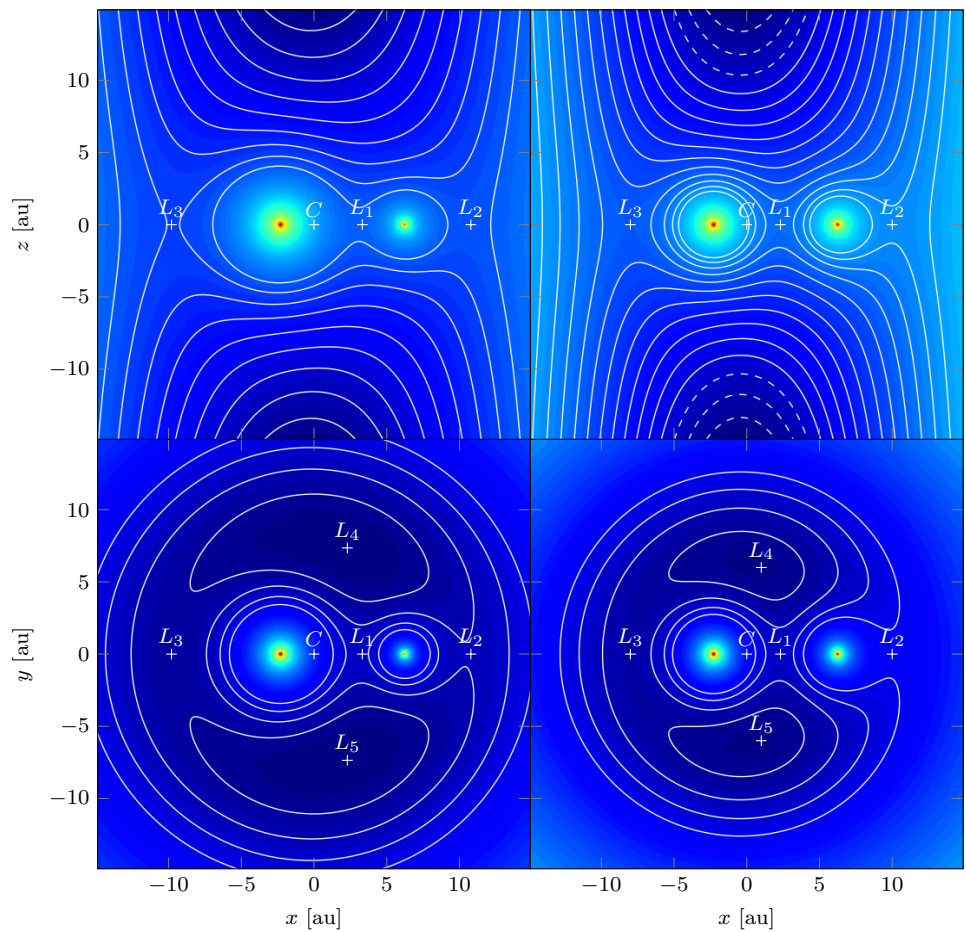

We modelled the effect of the radiative pressure reducing the gravity of
the primary component by a factor of $1-f_\mathrm{grav}$ to take into
account the acceleration mechanism  of the wind, where $f_\mathrm{grav}$
is the ratio of the radiation pressure force per unit mass to the
gravitational force per unit mass and is given by the expression
\citep[e.g.,][]{2003agbs.conf..461J}:

\[
  f_\mathrm{grav} = \frac{-\dif P}{\rho \dif r}\left(\frac{GM_1}{r_1^2}
  \right)^{-1},
\]
where symbols have the same meaning as in previous equations.  We assume
that the value of the $f_\mathrm{grav}$ factor is independent of
position \citep{2009A&A...507..891D} and use $f_\mathrm{grav} = 0.5$ in
our simulations.

Fig.~\ref{fig:chcyg} shows the Roche equipotential contours and the
location of the Lagrangian points in the CH Cyg system (see
Table~\ref{tbl:par}) when the radiation pressure of the AGB star is
ignored.  The shape of equipotential contours is modified when the
primary star is assumed to exert radiation pressure or pulsations that
drive the wind \citep{2009A&A...507..891D}.  We show the contours for a
ratio of the radiation pressure to gravitational force of 0.5 in the
right-hand column in Fig.~\ref{fig:chcyg}.  The surface where the wind is
accelerated is a substantial fraction of the Roche lobe radius of the
mass-losing primary star. This radius is about two to three times larger
than the stellar radius for most AGB stars \citep[see
e.g.,][]{2007ASPC..378..145H}.

The material in the disc around the WD is removed after each time step using
the expression from \citet{2002A&A...387..550G}:
\[
  \Delta \rho = \min\left(0,f_\mathrm{acc} \Delta t\right) \max\left(0,\rho -
  \rho_\mathrm{av} \right),
  \label{eq:acckley}
\]
where $f_\mathrm{acc}$ is a constant fraction of the order unity, $\Delta t$ denotes
the time-step and
$\rho_\mathrm{av}$ is the average density in
the region
$r_\mathrm{acc}<|\mathbf{r}-\mathbf{r}_2|<
2r_\mathrm{acc}$ following \citet{2008MNRAS.386..164P}.
The accreted
material on to the secondary powers the accretion luminosity. 
Radiative feedback can potentially change the flow and accretion rates
\citep{2004MNRAS.349..678E}, but we have neglected its effect in our
model.

\begin{table}
  \caption{CH Cyg component parameters used in the simulations from
    \citet{2009ApJ...692.1360H}.}
  \label{tbl:par}
  \centering
  \begin{tabular}{cS[table-format=>4.2]
		   s[table-unit-alignment=left]}
    \toprule
    Parameter            & \multicolumn{2}{c}{Observed values}      \\
    \midrule
      \multicolumn{3}{c}{Red giant}           \\
      $M_\mathrm{bol}$ & -4.5              \\
      $L$              & 5000                 & \lsun \\
      $T_\mathrm{eff}$ & 3100                 & \kelvin   \\
      $R$              & 280                  & \rsun \\
      $M$              & 2                    & \msun \\
      \midrule
      \multicolumn{3}{c}{White dwarf}         \\
      $L$ & 0.25                 & \lsun \\
      $M$ & >0.56                & \msun \\
      \midrule
      \multicolumn{3}{c}{Long-period orbit} \\
      $P$ & 15.6                 & \year        \\
      $a$ & 8.5                  & au        \\
      $i$ & 84                   & \degree   \\
      \midrule
      \multicolumn{3}{c}{Circumstellar shell} \\
      $R_\mathrm{inner}$ & 22                   & au        \\
    \bottomrule
  \end{tabular}
\end{table}

These numerical experiments were conducted using a dust acceleration
radius distance of $\sim$ \SI{2.7}{au}, where the wind is accelerated
slightly beyond the escape velocity by the absorption of stellar
radiation. For a typical value of the radius of the red giant of
\SI{300}{\rsun}, the dust acceleration occurs at about twice the size of
the stellar photosphere and inside the Roche lobe with the acceleration
zone being about a factor of 2 smaller than the Roche radius of the
system.  The wind velocity, density and pressure are set in all grid
cells within the wind acceleration radius of the primary component.
Although this method does not include the dust as a coupled component
and the full physics of the wind acceleration mechanism is not
considered, it can provide the expected range of terminal wind
velocities that are crucially important for deriving accretion rates on
the secondary.

The donor star is assumed to have radial pulsations in its outer
envelope, which are characteristic of AGB stars.  These pulsations are
considered as a source of the mass-loss fluctuations in the evolved
star. As this effect is relevant for our investigation of accretion rate
in symbiotic systems, we have included a pulsation period of
\SI{100}{\day} and an amplitude of \SI{1}{\km\per\s} in the material that
is introduced in the dust acceleration radius.  The amplitude of stellar
pulsations that can possibly produce the wind variability in the primary
can be constrained by models of evolved stars \citep[see
e.g.,][]{2007BaltA..16...26P}.

\section{Results and Discussion}
\label{sec:results}

In this section, we present the results of wind accretion simulations
using the hydrodynamical code described in Section~\ref{sec:model}.
Table~\ref{tbl:par} shows the fiducial parameters used in the accreting
wind model of the variable CH Cyg system studied by
\citet{2009ApJ...692.1360H}.

We parametrize the wind from the AGB star using the mass-loss, dust
acceleration radius and initial velocity for the wind.  Several
steady-state cases are compared using uniform and adaptive grids to
evaluate the accuracy of the results (cases 1--3).
Table~\ref{tbl:cases3d} shows the grid and wind parameters of different
simulations in the three-dimensional runs, including short outburst
simulations (cases 4--6).  The computational domain is divided into
squared blocks where $n_x$, $n_y$ and $n_z$ indicate the number of grid
points along the corresponding coordinate directions.  Based on our
previous work with two-dimensional calculations and more recent results,
an adaptive mesh grid provides an adequate method to resolve the
accretion region around the companion (\citealt{2007A&A...471.1043D};
\citetalias{2009ApJ...700.1148D}; \citealt{2013MNRAS.433..295H}). All
the simulations are run for approximately two orbital periods when the
system reaches a quasi-steady state.  As noted in the previous section
we compute the mass accretion rates on to the secondary normalized to
the mass-loss rates as a function of time during the simulation by
integrating the mass within a control volume.

In the following, we present the results from three-dimensional
simulations using a dynamically refined grid as they allow for adequate
grid resolution.  We consider a range of
parameters inferred from the observations of CH Cyg summarized
in Table~\ref{tbl:par} \citep{2009ApJ...692.1360H} and vary the wind
velocity, mass-loss  and wind acceleration radius in our simulations
to account for realistic outburst events.
We adopt the most probable solution of the orbit, with an orbital period
of \SI{15.58}{\year} and masses of \SI{2}{\msun} and \SI{0.6}{\msun} for our
model.

\begin{figure*}
  \centering
  \begin{tabular}{cc}
  \includegraphics[width=\columnwidth]{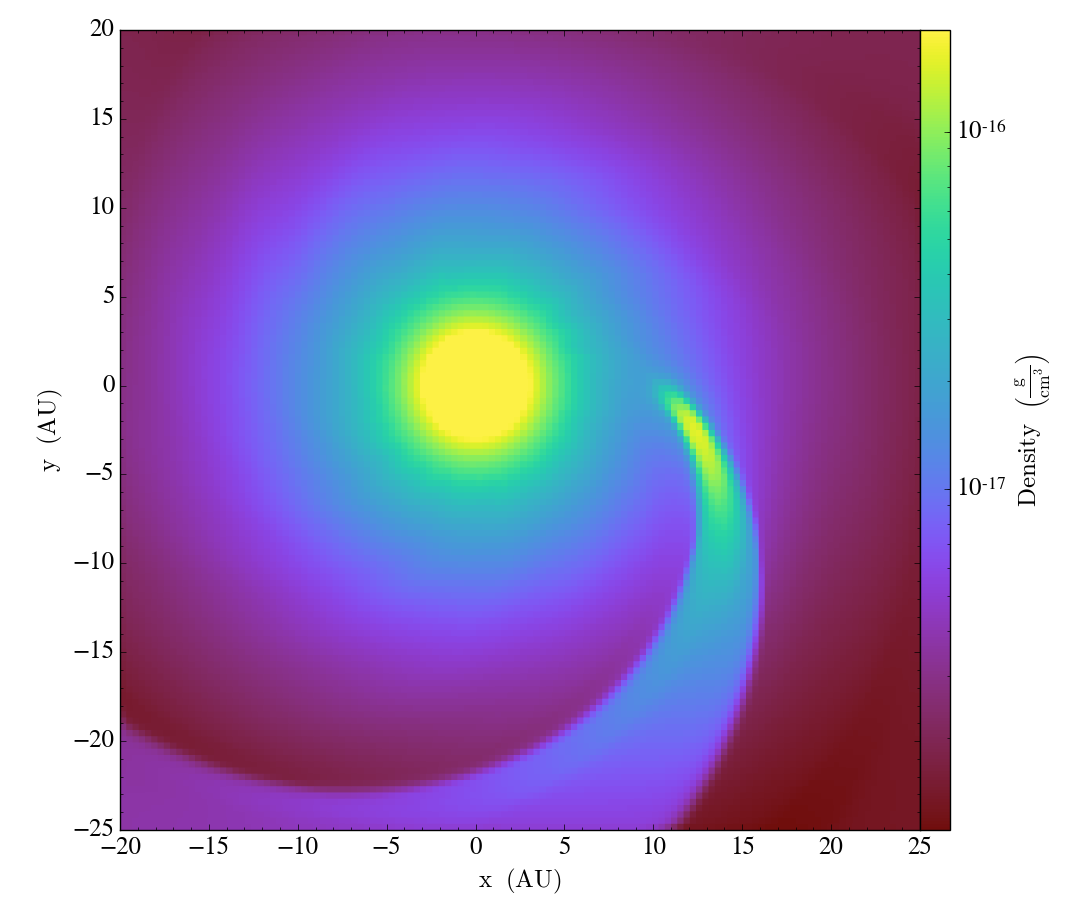} &
  \includegraphics[width=\columnwidth]{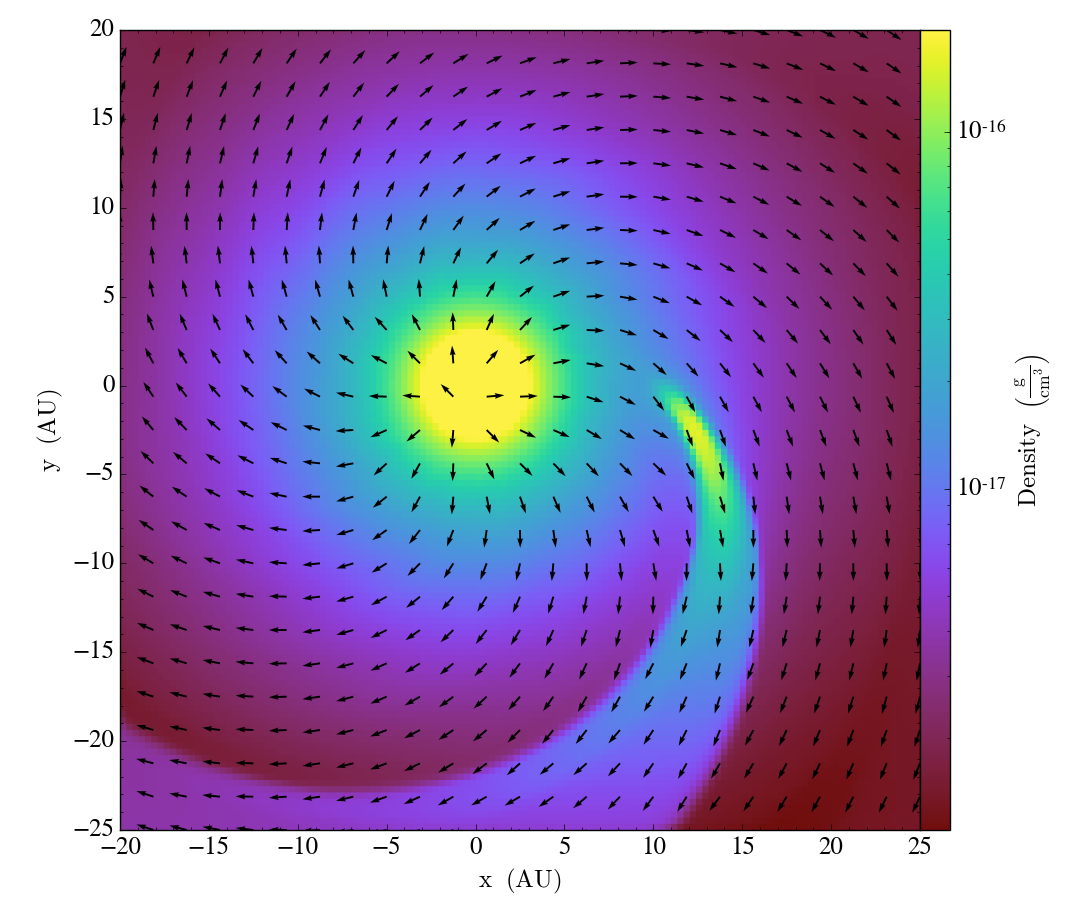} \\
  \includegraphics[width=\columnwidth]{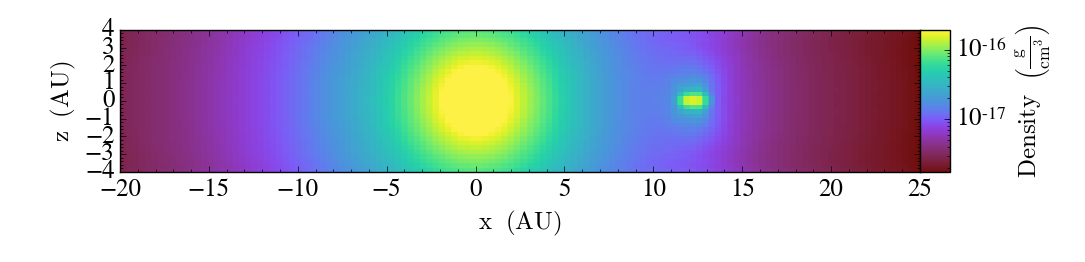} &
  \includegraphics[width=\columnwidth]{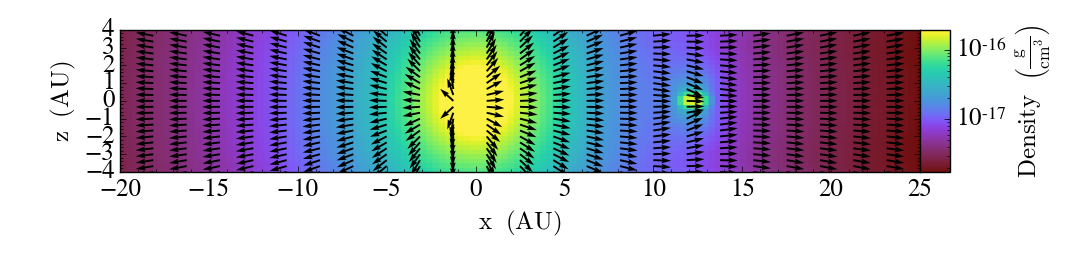}
  \end{tabular}
  \caption{Snapshot of gas density contours in logarithmic scale in the
    orbital plane slice (upper left-hand panel) and perpendicular plane
    slice (lower left-hand panel)
    through the computational domain taken after one orbit for the
    case 1 simulation described in Table~\ref{tbl:cases3d} after one
    orbital period.
    The velocity field is shown in the orbital plane slice (upper
    right-hand panel) and perpendicular plane (lower right-hand panel)
    slices for the case 1 simulation.  The velocity is calculated in the
    non-inertial frame and has been normalized by the spatial distance
    for clarity.  The mass-losing star is located on the origin of the
    $x$--$y$ plane and the accreting component is located at $(x,y) = (8.5,
      0)$.
  }
  \label{fig:3dcont}
\end{figure*}

\subsection{Steady-wind simulations}
\label{sec:steady}

We cover the range of \SIrange{20}{50}{\km\per\s} of the initial wind
velocity at the dust acceleration surface for the steady-state wind
simulations.  Initially, we perform a comparison of flow structures in the
three-dimensional simulations with the two-dimensional results described
in \citetalias{2009ApJ...700.1148D}.  Our choice of wind velocities
ensures that most of the wind material that is injected at the dust
acceleration boundary will escape from the gravitational pull of the
binary system and leave the computational domain.  For this choice of
wind parameters, the flow has a Mach number \textit{M} = 2--4 at the
position of the accretor.

\begin{figure}
  \centering
  \includegraphics[width=\columnwidth]{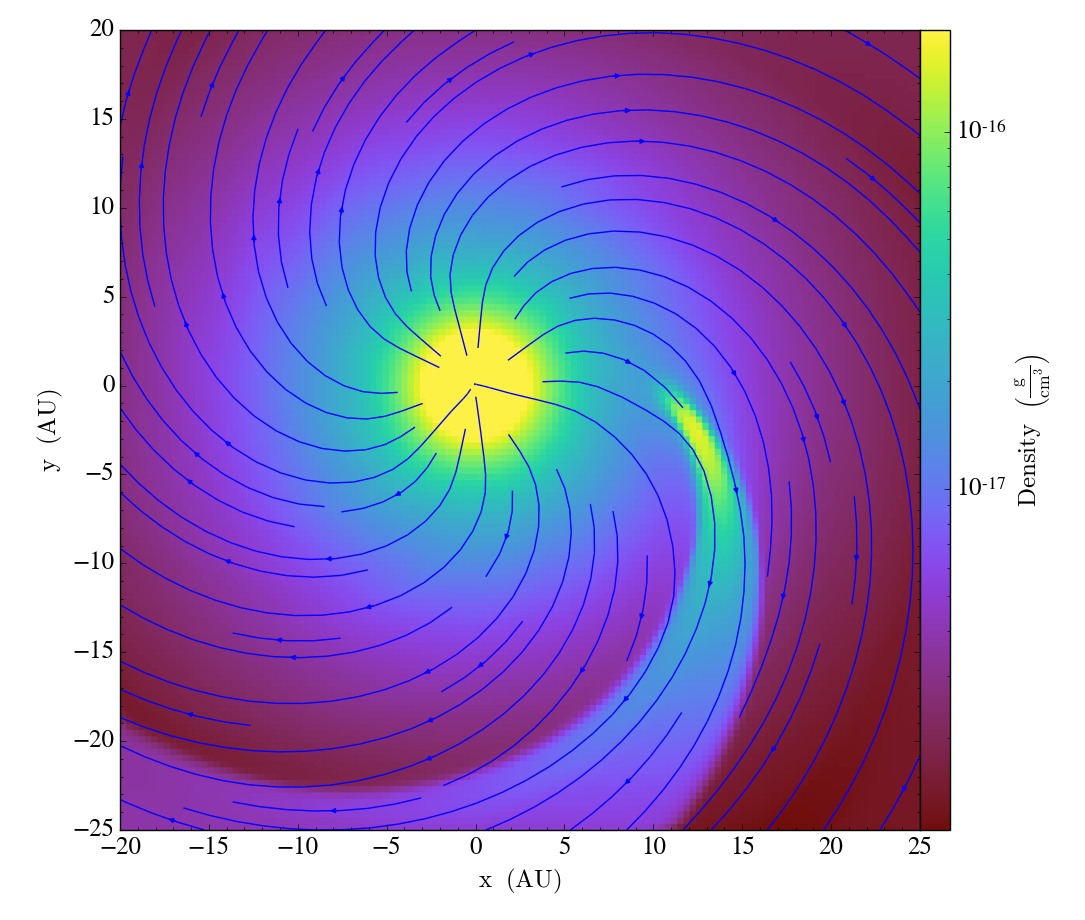}
  \caption{Streamlines in the orbital plane
    in the non-inertial frame
    for the case 1 simulation in Table~\ref{tbl:cases3d}.
  }
  \label{fig:3dstream}
\end{figure}

We show the density distribution of the perturbed wind in the simulation
with a terminal wind speed of \SI{20}{\km\per\s} in
Fig.~\ref{fig:3dcont} after one orbit. The wind material forms a spiral
shock around the secondary that is detached from the accretor and is
focused on the orbital plane, as has been seen in previous numerical
works \citep{2008ApJ...675L.101E,2016MNRAS.455.4351P}.  A flattened elongated
overdensity at the position of the secondary forms by the accretion of
shocked material from the stellar wind.  The velocity field geometry in
a quasi-steady state after one orbital period is shown in the right-hand
panels of Fig.~\ref{fig:3dcont}.  In addition, we show the orbital velocity
of the companion added to the total velocity of the gas. Some wind
material from the giant star that passes close to the secondary and is
shocked by the spiral arm has an escape velocity smaller than that of the
companion.  This material is accreted along the accretion line behind
the companion.  The gas stream observed in previous two-dimensional
simulations and SPH models is considerably weaker
\citep[\citetalias{2009ApJ...700.1148D};][]{2012BaltA..21...88M}.
The deflection of the streamlines in the non-inertial frame, which are
defined as the lines that are always tangential to the velocity, is
shown in Fig.~\ref{fig:3dstream}.

Since we have used the reduced gravity expression described in
Section~\ref{sec:model}, the  focused wind accretion in this system
shows a complex dynamics, resulting in arc structures and low-density
cavities that form in the envelope beyond the accreting companion.  We
observe that a significant fraction of the material in the wind lost by
the giant component is focused on the orbital plane and partially
accreted on to the hot compact star.  Additionally, a bow shock and a
spiral-like tail structure forms near the position of the secondary and
on the farther side from the primary  due to the supersonic orbital
motion creating an asymmetric configuration.  The shock winds around the
primary and some of the material accumulates in the region around the
secondary, forming an accretion disc.  This structure is similar to that
observed in other numerical simulations of wind accretion in binary
systems
\citep[e.g.,][]{1993MNRAS.265..946T,1999ApJ...523..357M,2008ApJ...675L.101E}.
A small fraction of the matter in the stellar wind falls back on to the
primary after being deflected by the companion.

\begin{figure}
  \begin{tikzpicture}
    \begin{axis}[
      xlabel={time ($P$)},
      ylabel={$f \times \num{e2}$},
      xmin=0,
      xmax=2,
      mark size=1pt,
      ]
      \addplot[only marks,green] table {accretion.txt};
      \addplot[mark=none, dashed, samples=2, domain=0:2, black] {9.4};
    \end{axis}
  \end{tikzpicture}
  \caption{Time evolution of the accretion ratio, defined as the
    accretion rate divided by the mass-loss rate from the primary star.
    The dashed line shows the accretion rate predicted by BHL\@.
    See the text for the details of how the accreted mass by the
    companion is computed.
  }
  \label{fig:accretion}
\end{figure}
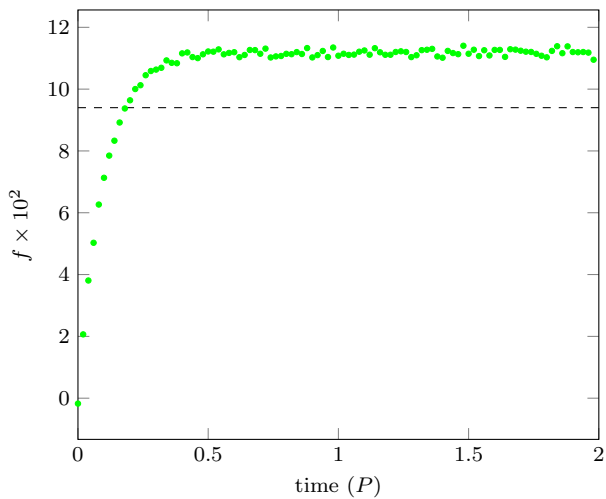

We define the accretion ratio as the accreted mass over the mass lost
by the secondary integrated over a hundredth of an orbit:
\[
  f = \frac{\dot{M}_\mathrm{acc}}{\dot{M_1}}.
\]
For comparison, we use the BHL accretion rate calculated from numerical
solutions of BHL flow by \citet{1985MNRAS.217..367S}, as referred to in
equation~(32) of \citet{2004NewAR..48..843E}. The estimated mass-capture rate
by the secondary in the numerical approach is:
\begin{equation}
  \dot{M}_\mathrm{BHL} = \frac{4 \pi G^2 M_\mathrm{2}^2
  \rho_\infty}{(c_s^2 + v_\infty^2)^{3/2}},
  \label{eq:hl}
\end{equation}
where $c_s$ is the sound speed of the gas; other symbols have their usual
meaning.  Note that this estimate includes an extra factor of 2 based
on two-dimensional numerical simulations
\citep[e.g.,][]{1985MNRAS.217..367S}.
In this approximation, the fluid is supposed to be
non-interacting except on the downstream side of the shock where
material flows on to the accretor. This framework was originally
applied to the flow around single star moving through the interstellar
medium.  However, the accretion rate given in equation~(\ref{eq:hl}) is
frequently used as a benchmark with numerical simulations of accretion
in isolated stars and binary systems \citep[see
e.g.,][]{2004NewAR..48..843E,2005A&A...434...41E}.
Fig.~\ref{fig:accretion} shows the time evolution of the
accretion ratio $f$ compared with the expected BHL accretion rate for
this system.  After a brief transient period, the accretion ratio reaches
a steady value about 20\% higher than the BHL estimate defined by
equation~(\ref{eq:hl}).  The material transferred from the donor to the
accretor carries angular momentum that will introduce a change in the
orbital parameters with a decreasing orbital period. However, this
change occurs on a time-scale much greater than the simulation
characteristic time.

\subsection{Outburst simulations}

\begin{table*}
  \begin{threeparttable}
  \caption{Parameters of the three-dimensional simulations with a steady
    flow and wind outburst on the dust formation radius after the flow
    approaches a steady state.  The computational domain is centred on
    the primary covering (300, 300, 10) au in the ($x, y, z$)
    coordinates. The outburst start time is 0.5 orbits, and the duration
    of the outburst is 0.1 orbits for simulations 4--6.
  }
  \label{tbl:cases3d}
    \begin{tabular}{Sccc
		    S[table-format = 2]
		    S[table-format=1e+1,
		    table-number-alignment = center]
		    S[table-format = 2]
		    S[table-format=1e+1,
		    table-number-alignment = center]}
      \toprule
	{Case} & Grid & Resolution & Refine\tnote{\emph{a}} &
	  {$v_\infty$\tnote{\emph{b}}} &
	  {$\dot{M}$\tnote{\emph{c}}} &
	  {$v_{\infty,\mathrm{burst}}$\tnote{\emph{d}}} &
	  {$\dot{M}_\mathrm{burst}$\tnote{\emph{e}}}\\
	& & ($n_x \times n_y \times n_z$) & &
	{(\si{\km\per\s})} & {(\si{\msun\per\year})} &
	{(\si{\km\per\s})} & {(\si{\msun\per\year})} \\
      \midrule
	1 & Uniform & $320 \times 320 \times 128$ &   & 20 & e-6  \\
	2 & AMR     & $320 \times 320 \times 128$ & 3 & 50 & e-6  \\
	3 & AMR     & $320 \times 320 \times 128$ & 3 & 20 & 2e-6 \\
      \midrule
	4 & AMR & $512 \times 512 \times 32$ & 5 & 20 & e-6 & 80  & e-6  \\
	5 & AMR & $512 \times 512 \times 32$ & 5 & 20 & e-6 & 20  & 2e-6 \\
	6 & AMR & $512 \times 512 \times 32$ & 5 & 20 & e-6 & 200 & 2e-6 \\
      \bottomrule
    \end{tabular}
    \begin{tablenotes}
      \item [(\emph{a})] Number of AMR refinement levels.
      \item [(\emph{b})] Terminal wind velocity.
      \item [(\emph{c})] Mass-loss rate from the primary.
      \item [(\emph{d})] Terminal wind velocity during outburst.
      \item [(\emph{e})] Mass-loss rate from the primary during outburst.
    \end{tablenotes}
  \end{threeparttable}
\end{table*}

We performed three sets of numerical experiments introducing a stellar
outburst in the mass-losing star during a fraction of an orbital period
to study the effect of the variation in the wind properties on the
envelope and large-scale spiral structure.  A smooth variation of the
wind properties is included during a short fraction of an orbital period
to prevent numerical instabilities.  An estimate of the changes in the
mass accretion ratios is of special interest since it is expected in
various types of interacting binaries.  This variation produces a smooth
spherically symmetric shell that is focused towards the equatorial plane
of the system by the gravity of the companion.  The size of the physical
domain was increased by a factor of 6 in the $x$-and $y$-directions, and
we used two additional mesh refinement levels as compared with the
steady-wind simulations presented in Section~\ref{sec:steady} to improve
the resolution around the secondary and bow shocks.  The parameters of
these simulations are summarized in Table~\ref{tbl:cases3d}.  The values
of the wind parameters during the eruption were varied over several
time-steps to avoid stability problems in the simulation.

\begin{figure}
  \centering
  \includegraphics[width=\columnwidth]{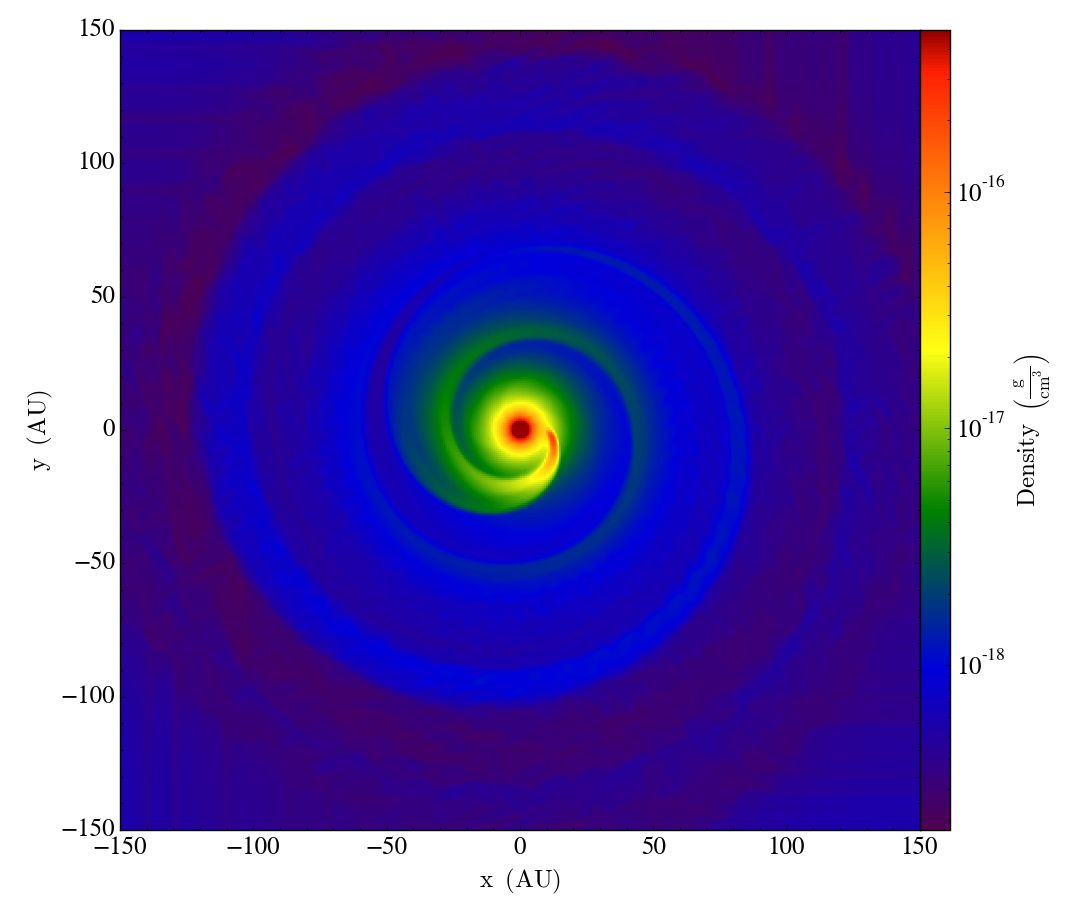}
  \caption{Density contour in logarithmic scale in the orbital plane
    for the case 3 simulation in Table~\ref{tbl:cases3d} after
    introducing the perturbation.
  }
  \label{fig:3dwide}
\end{figure}

We first consider a system with the orbital parameters of CH Cyg and a
mass-loss of \SI{e-6}{\msun\per\year} from the primary star and a
terminal wind speed of \SI{20}{\km\per\s}.  Thus, the wind speed and the
orbital velocity have comparable values.  This speed ensures that the
wind can escape from the computational domain without falling back on
the primary.  After the flow reaches the quasi-steady-state solution,
spiral shocks form, surrounding the secondary from the stellar wind
material of the giant star.  The wind velocity at the dust acceleration
distance is increased by a factor of 4 during an interval of 0.1 orbits
and then brought back to its initial value (case 4 in
Table~\ref{tbl:cases3d}).  The radial speed is modified for all grid
cells between two radii $0.2a$ and $0.3a$ around the primary star.
During the burst the wind leaves the wind acceleration surface in a
highly supersonic regime.  Thus the perturbation takes about 0.04 orbits
to reach the location of the accretor. The flow produced by the burst is
deflected by the secondary and produces a temporary increase in the
temperature.  This pulse generates an expanding shell that modifies
considerably the structure of the accretion flow and the rotating disc
structure observed around the accretor during the quiescent phase.  When
the shell leaves the boundaries of the domain the accretion flow returns
to its initial equilibrium state with a bow shock spiralling around the
system.  The mass accretion ratio is decreased during the outburst
period although it is greater than the BHL prediction from
equation~(\ref{eq:hl}). In Fig.~\ref{fig:3dwide} we show the density in the
orbital plane.

\begin{figure}
  \begin{tikzpicture}
    \begin{axis}[
      xlabel={time ($P$)},
      ylabel={$f \times \num{e2}$},
      xmin=0,
      xmax=2,
      mark size=1pt,
      ]
      \addplot[only marks,red] table {accretion_burst.txt};
      \addplot[only marks,green] table {accretion_burst2.txt};
      \addplot[mark=none, dashed, samples=2, domain=0:2, black] {9.4};
    \end{axis}
  \end{tikzpicture}
  \caption{The time evolution of the accretion ratio relative to the
    background mass-loss rate for an outburst simulation where the
    outflow velocity is modified to \SI{80}{\km\per\s} (green points)
    and where the mass-loss rate of the primary component is increased
    to \SI{2e6}{\msun\per\year} (red points) after 0.6 orbits .  The
    dashed line shows the accretion rate predicted by BHL in the
    quiescent phase.  See the text for the details of how the accreted
    mass by the companion is computed.}
  \label{fig:accretion_burst}
\end{figure}
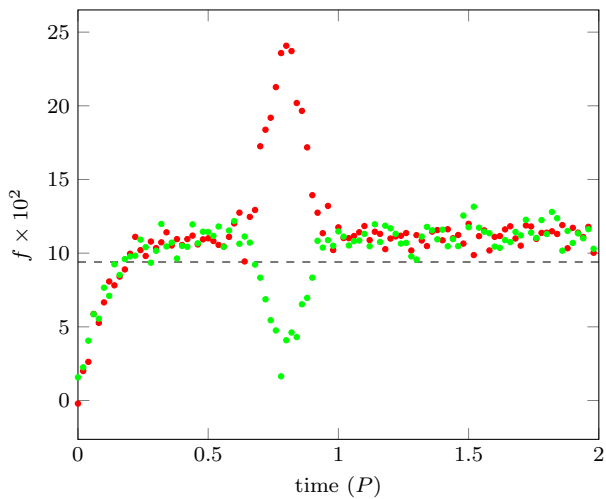

An important indication of observational signatures of the outburst
phase is the amount of matter that is accreted by the secondary and
their temporal behaviour.  We show the time evolution of the accretion
ratio for simulations 4 and 5 in Fig.~\ref{fig:accretion_burst} compared
with the standard BHL planar accretion.  This approximation, where the
effects of pressure and non-parallel flow are ignored, provides a good
estimate, but the observed accretion rates in the simulations are, on
average, higher.  The accretion ratios are modified during the outburst
phase and settle to a steady-state value during the period when the
flow becomes stationary once again.  In this model, the accretion rate
from quiescence is increased by about 40\% compared to the BHL value
with some variability, which could be explained by the limited resolution
around the secondary and numerical noise.

\begin{figure*}
  \centering
  \begin{tabular}{cc}
    \includegraphics[width=\columnwidth]{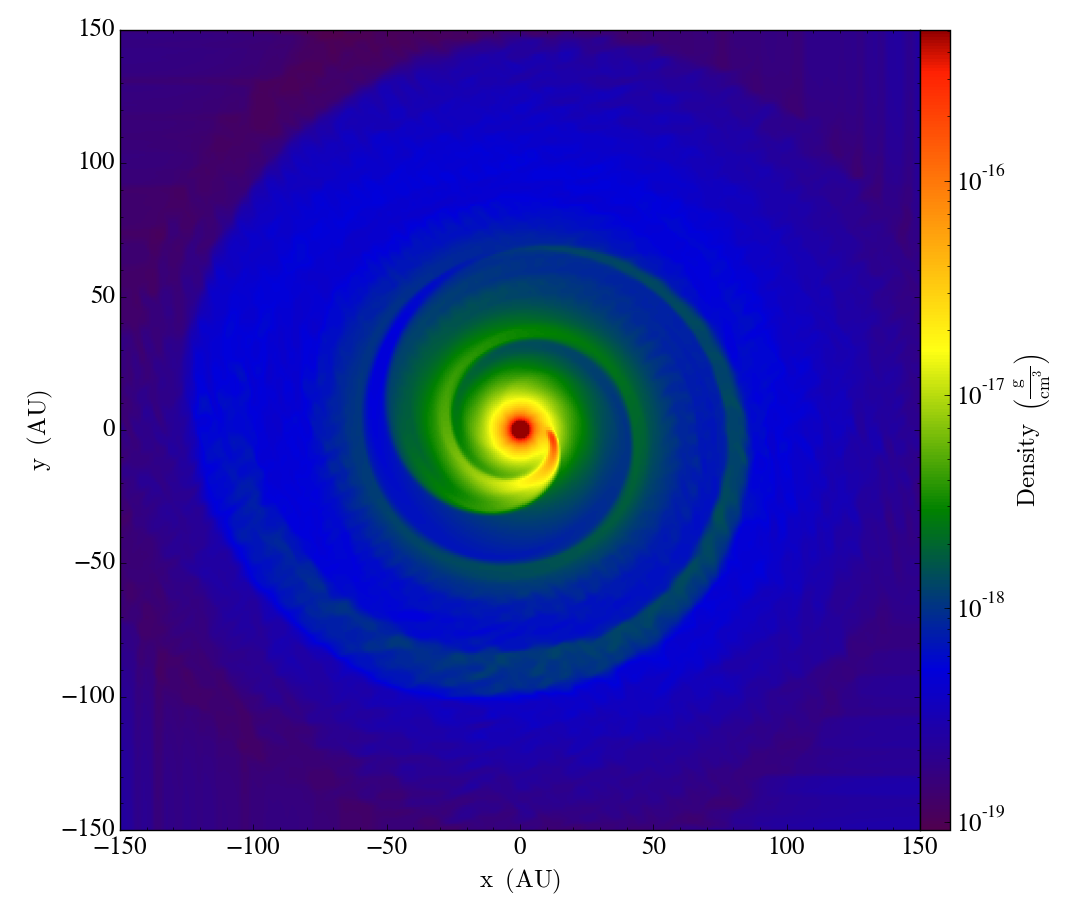} &
    \includegraphics[width=\columnwidth]{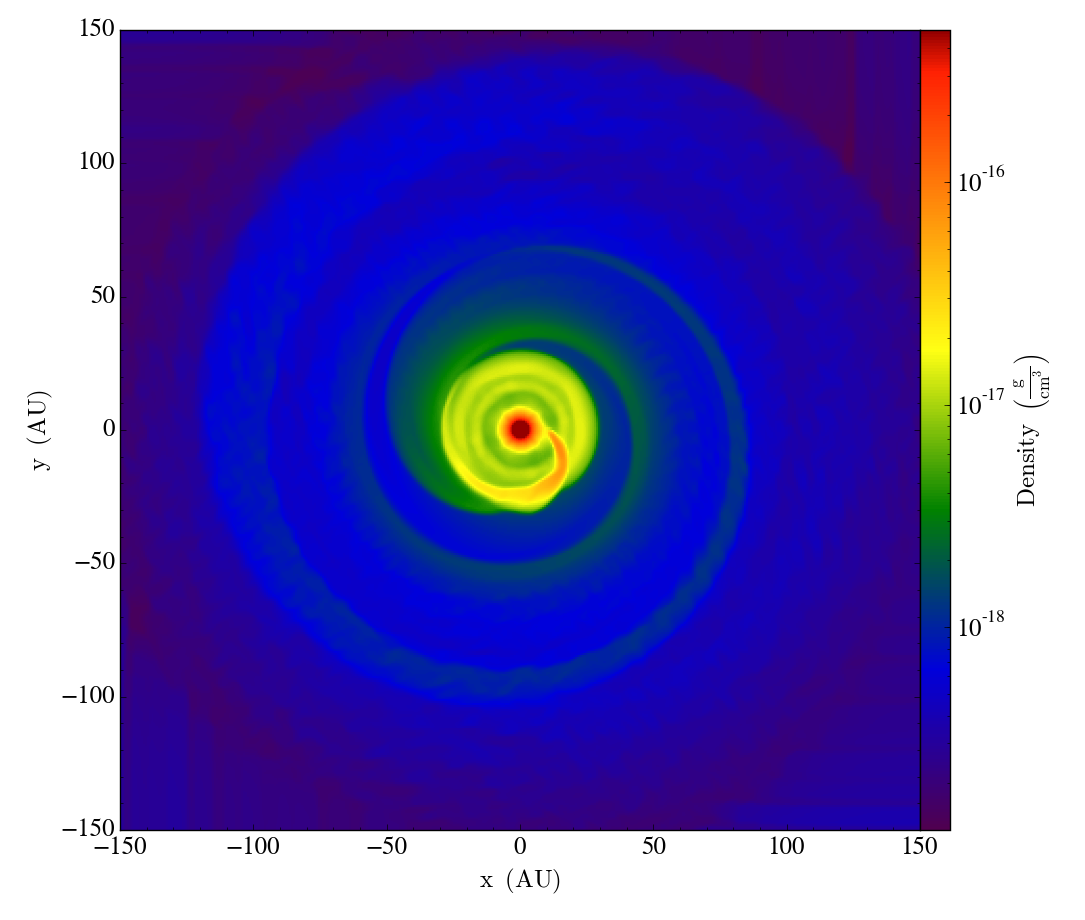}\\
    \includegraphics[width=\columnwidth]{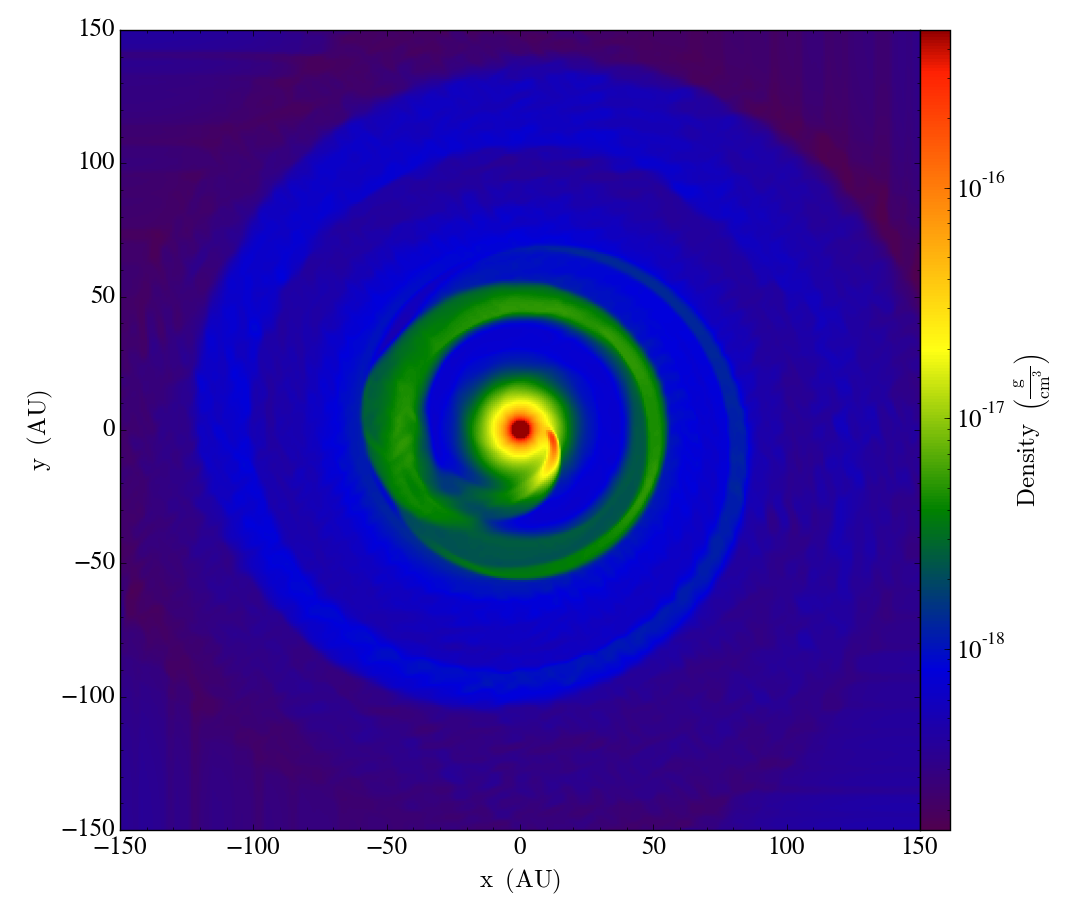} &
    \includegraphics[width=\columnwidth]{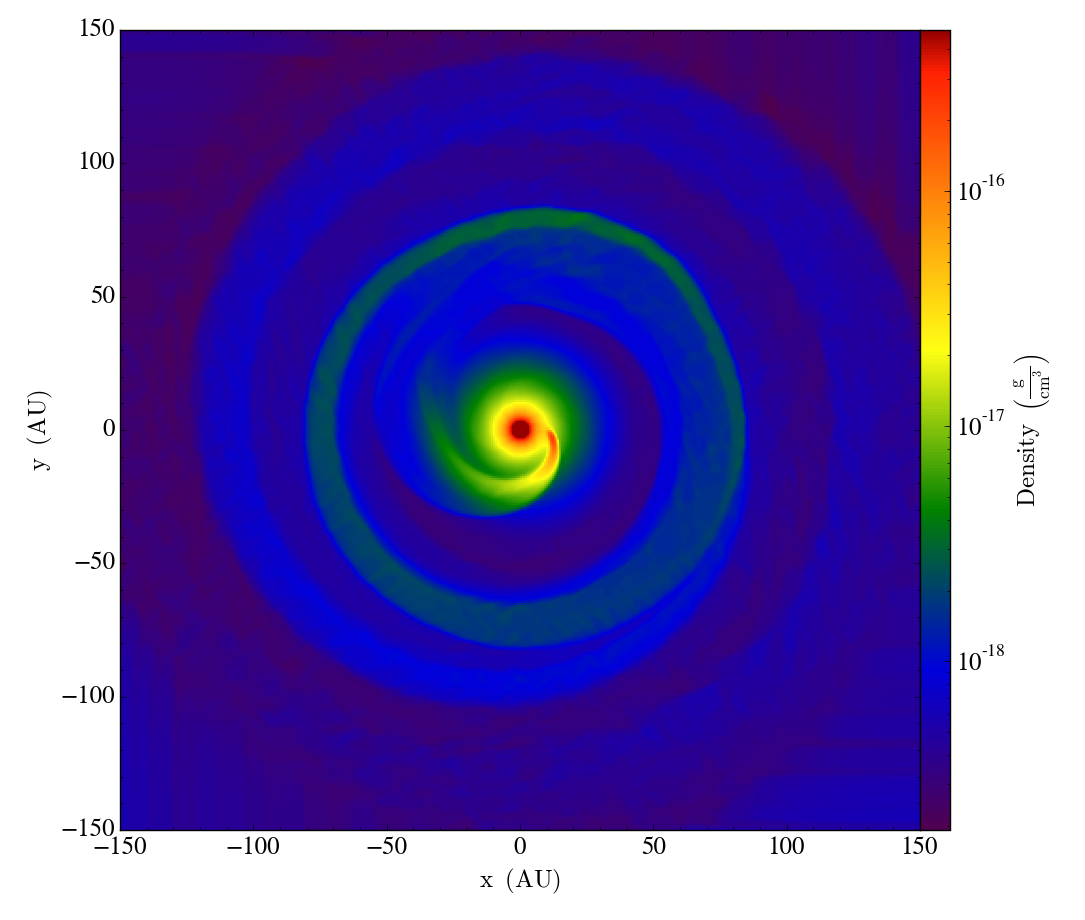}
  \end{tabular}
  \caption{Sequence of density contour snapshots in logarithmic
    scale in the orbital plane for the case 5 simulation in
    Table~\ref{tbl:cases3d} at 0.5, 0.7, 0.8 and 1 orbit.
  }
  \label{fig:burst}
\end{figure*}

We ran another numerical experiment with an outburst phase starting from
a steady-state solution for a scenario where the wind velocity was kept
constant and the mass-loss rate increased by a factor of 2 during
0.1 orbit and then returned to its initial value of \SI{e-6}{\msun\per\year}
(simulation 5 in Table~\ref{tbl:cases3d}). This simulation was run for
further 1.5 orbits after introducing the disturbance to consider the
settling period of the accretion flow.  The spiral structure was
substantially perturbed by the outburst, and the spiral arms merged
together forming a ring-shaped surface. The effect of the burst
introduced in the mass-loss rate disappeared after two additional orbital
periods. A sequence of density snapshots showing the temporal evolution
of the flow patterns in the orbital plane is shown in
Fig.~\ref{fig:burst}.  The flow pattern is significantly more
complicated than in the constant mass-loss and wind speed simulations
described previously.

Finally, we carried out a simulation where the wind velocity was increased
by an order of magnitude for one-tenth of an orbit and the mass-loss
rate was kept constant with a value of \SI{e-6}{\msun\per\year}
(simulation 6 in Table~\ref{tbl:cases3d}).  With a larger velocity
increase, the flow shows a more extended disturbance of the steady flow
than in the previously studied cases.  After about 0.2 orbit from the
start of the disturbance, the flow solution returns to the steady state
obtained in the other simulations.  In all simulations of the outburst
phase, the accretion ratio varies over a similar time over which the wind
parameters are modified.

Despite the oscillations in the wake during the outburst phase, the
accretion flow in the simulations is steady and there is no evidence of
a violent instability, such as the flip-flop behaviour that has been
previously hinted at in some numerical simulations of BHL flow
\citep{2013ApJ...767..135B}. The large-scale structure of the wake stays
undisrupted. These results are consistent  with previous high-resolution
three-dimensional simulations of BHL accretion
\citep{2005ARep...49..884M,2012ApJ...752...30B}. Thus, introducing these
perturbations is not sufficient to affect the stability of the flow and
accretion, and does not bring unstable behaviour
\citep{1987MNRAS.226..785M}.

\section{Summary and Conclusions}
\label{sec:discuss}

In this paper, we have presented a three-dimensional numerical model of
focused wind accretion with an adiabatic equation of state in the
context of interacting binary systems, for example, symbiotic systems such as
the variable CH Cyg, and considered the dynamical aspects of the
accretion flow in the simulations.  A strong spiral shock forms, which
is caused by the gravity of the accretor with a moderately wide opening angle.
The flow pattern is similar to the two-dimensional planar geometry studied
in \citetalias{2009ApJ...700.1148D}.  However, there are significant
differences, including the development of a larger asymmetry in the inner
and outer spiral shocks.  Our numerical method does not need artificial
viscosity to stabilize the flow and is well suited for studying a high
Mach number flow around the accretor.  The obtained wind solutions and
mass accretion rates suggest that the accretion via the wind from the
giant star on to the companion occurs in a more efficient manner by a
factor of 1.5 than in the standard BHL estimate based on numerical
simulations \citep[e.g.,][]{2004NewAR..48..843E}.

We have performed several sets of simulations of accretion flow in
symbiotic binary systems with varying mass-loss rates and wind speeds
during a fraction of an orbit.  First,  we increased the wind speed at
infinity by a factor of 4. The second run used the original wind
speed and doubled the mass-loss rate from the donor.  In the last
scenario, we increased both the wind speed and mass-loss rate.  The
first and third experiments show a change in the accretion rates on to
the secondary of around a factor 2 larger than the BHL accretion rates
and a largely disturbed flow as the outburst material interacts with the
accretion wake.

For the velocity outburst simulation, the accretion ratio decreases
compared with the quiescent state. However, the accretion is higher than
the value expected from the BHL estimate also during the transient
outburst event.  The variation in the accretion rate is consistent
within a factor of 2 with the steady-state accretion expected from
equation~(\ref{eq:hl}), which is substantially larger than the value for the
pressure-free flow originally studied by \citet{1939PCPS...35..405H}.
None the less the simulations where we vary the mass-loss rate match more
closely the BHL approximation given by equation~(\ref{eq:hl}).  The time-scales
of the variations introduced in the mass-loss rates are of the order of
weeks to months, which could be associated with fluctuations in the
brightness of these objects \citep{2000A&AS..146..407B}.  These results
agree well with two- and three-dimensional numerical simulations of BHL
accretion \citep{1994ApJ...427..351R,2005A&A...434...41E}.  We do not
observe any instability in the accretion column that has been described
in previous works \citep{1990ApJ...358..545S,2005A&A...434...41E}, as
expected for the Mach number of the flow that we have considered.  In
addition, there is no oscillation of the disc rotation in the
simulations such as observed in the flip-flop instability
\citep[e.g.,][]{2000MNRAS.313..198P,1999A&A...346..861R}, in accord with
previous three-dimensional simulations in spherical coordinates
\citep{2012ApJ...752...30B}.

Understanding the origin of the outflows and jets in wind interacting
systems is crucial for providing insights into the mass-transfer mechanism.  A
limitation of the models presented here is that the wind acceleration
mechanism is prescribed without including the dust formation and
radiative pressure that are responsible for the wind formation.  Such
numerical analysis may bring new effects as a treatment of dust-driven
acceleration in the wind model may modify the outflow formation and
mass-loss rate.  We plan to address these effects in future works with
an application to the interpretation of CH Cyg and other symbiotic systems.

\section*{Acknowledgements}

This work was supported by NASA's Planetary Astronomy Program.
MK acknowledges support provided by NASA grants HST
GO-NAS12761, and Chandra GO02-13031, and support from the Chandra X-ray
Center operated by the Smithsonian Astrophysical Observatory under NASA
Contract NAS8-03060.  We thank the anonymous referee for carefully
reading our paper and providing constructive comments.  The numerical
codes used in this article are based on the open source \piernik{} code,
and the publicly available \flash{} code, which is in part developed by
the DOE NNSA-ASC OASCR Flash Center at the University of Chicago.
We especially thank Artur Gawryszczak and Kacper Kowalik for many useful
discussions on the \piernik{} code.  All the analysis and visualization
of the data were carried out using the \textsc{yt} toolset by
\citet{2011ApJS..192....9T}\footnote{The package is available for
download from \url{http://yt-project.org}.} and the \texttt{pynbody}
package \citep{pynbody}\footnote{The package is available at
\url{https://github.com/pynbody/pynbody}}.  The simulations presented in
this work were performed on computational resources supported by the
Princeton Institute for Computational Science and Engineering (PICSciE)
and the Office of Information Technology's High Performance Computing
Center and Visualization Laboratory at Princeton University.  This
research has made use of NASA's Astrophysics Data System.

\bibliographystyle{mnras}
\bibliography{ads,preprints,chcyg}

\bsp

\label{lastpage}

\end{document}